# A Comparison of Aggregation Methods for Probabilistic Forecasts of COVID-19 Mortality in the United States

Kathryn S. Taylor[1*], James W. Taylor[2]

[1]Nuffield Department of Primary Care Health Sciences, University of Oxford

[2]Saïd Business School, University of Oxford

16 August 2020

**Abstract**

The COVID-19 pandemic has placed forecasting models at the forefront of health policy making. Predictions of mortality and hospitalization help governments meet planning and resource allocation challenges. In this paper, we consider the weekly forecasting of the cumulative mortality due to COVID-19 at the national and state level in the U.S. Optimal decision-making requires a forecast of a probability distribution, rather than just a single point forecast. Interval forecasts are also important, as they can support decision making and provide situational awareness. We consider the case where probabilistic forecasts have been provided by multiple forecasting teams, and we aggregate the forecasts to extract the wisdom of the crowd. With only limited information available regarding the historical accuracy of the forecasting teams, we consider aggregation (i.e. combining) methods that do not rely on a record of past accuracy. In this empirical paper, we evaluate the accuracy of aggregation methods that have been previously proposed for interval forecasts and predictions of probability distributions. These include the use of the simple average, the median, and trimming methods, which enable robust estimation and allow the aggregate forecast to reduce the impact of a tendency for the forecasting teams to be under- or overconfident. We use data that has been made publicly available from the COVID-19 Forecast Hub. While the simple average performed well for the high mortality series, we obtained greater accuracy using the median and certain trimming methods for the low and medium mortality series. It will be interesting to see if this remains the case as the pandemic evolves.

[*] Email for correspondence: kathryn.taylor@phc.ox.ac.uk

## 1. Introduction

The coronavirus 2019 (COVID-19) was declared a pandemic by the World Health Organization on 11 March 2020 (WHO 2020a), and five months later, over 21 million individuals had been infected and COVID-19 has caused over 760,000 deaths worldwide (WHO 2020b). The COVID-19 pandemic has created enormous planning and resource allocation challenges. Governments are relying upon predictions of the numbers of COVID-19 cases, people hospitalized and deaths to help decide what actions to take (Adam 2020; Phelan et al. 2020). In this paper, we consider the short-term forecasting of reported deaths from COVID-19.

Forecasting methods are well established in providing predictions of uncertain events to decision makers across a variety of settings, ranging from energy providers and individuals relying on the weather outlook, to investors eager to gain insight into future economic conditions. Epidemiological forecasting models have been applied to both vector-borne diseases, including Dengue disease (Shi et al. 2016) and the Zika virus (Kobres et al. 2019), and contagious infectious diseases. These include the Severe Acute Respiratory Syndrome (SARS) (Ng et al. 2003), Ebola (Viboud et al. 2017) and the Middle East respiratory syndrome (MERS) (Da'ar et al. 2018). Numerous COVID-19 models have emerged (Adam 2020, COVID-19 Forecast Hub 2020). These models are based on different models and make different assumptions and therefore answer different questions (Holmdahl and Buckee 2020). Due to the lack of data, assumptions have to be made about several factors including the extent of immunity, transmission among people who are asymptomatic and how the public will react to new government restrictions. Paucity of data is a common challenge in forecasting infectious diseases (Lauer et al. 2020). Policy makers need to be aware of the limitations of the models, and need to be conscious of the uncertainty in predictions from these models (Sridhar and Majumder 2020).

Gneiting and Katzfuss (2014) describe how optimal decision-making relies on the availability of a forecast of a probability distribution, rather than just a single point forecast (see, for example, Gianfreda and Bunn 2018; Wang and Yao 2020). Interval forecasts are also valuable, as they can support real-time decision making and provide situational awareness (see, for example, Chen and Xiao 2012; Grushka-Cockayne and Jose 2020; Bracher et al. 2020). In this paper, we consider the case where probabilistic forecasts are provided by multiple forecasters for the cumulative total of deaths due to COVID-19. We aggregate the forecasts to extract the wisdom of the crowd. The benefit of collective decision making was highlighted by Galton in his seminal paper, as he reported the success of a crowd at a country fair in guessing the weight of an ox (Galton 1907). Aggregation provides a pragmatic approach to synthesizing the information underlying different forecasting methods. It also enables diversification of the risk inherent in selecting a single forecaster who may turn out to be poor, and it offsets statistical bias associated with individual forecasters who tend to be under- or overconfident. Forecast aggregation is often referred to as forecast *combining*.

Since the early work by Bates and Granger (1969) on combining point forecasts, methodology has evolved with the introduction of a variety of simplistic and sophisticated combining methods. Recent work has involved combining probabilistic forecasts. Winkler et al. (2019) predicts that probabilistic forecast aggregation will become more common due to developments in the field, the rising popularity of forecasting competitions, and raised awareness by increased reporting of probabilistic predictions in the media.

The aggregation of individual forecasts could be weighted in some way if a record of past accuracy is available for each individual forecaster. However, such a record is often not available, and, indeed, this is the case for our application, which involves the forecasting of cumulative deaths during the relatively early stages of a pandemic. Even as time passes, comparable historical accuracy will not be available for all forecasters, as their involvement with the forecasting exercise commenced at different times. A further concern is that the relative accuracy of forecasters may not be stable over time (Winkler et al. 2019). In this paper, we consider aggregation methods that do not rely on a record of past accuracy for each forecaster.

The success of the simple average for combining point forecasting has motivated its use for probabilistic forecasts. The median and trimmed means have also been proposed for forecasts of probability distributions and interval forecasts, as they provide simple, robust alternatives to the mean (Hora et al. 2013; Jose et al. 2014; Gaba et al. 2017). In this empirical paper, we evaluate the usefulness of these existing aggregation methods, applied to multiple probabilistic forecasts of COVID-19 mortality at the state and national level in the U.S, using data made publicly available (see COVID-19 Forecast Hub 2020). In Section 2, we describe the dataset and the rise in mortality due to COVID-19 in the U.S. We consider interval forecast aggregation in Section 3, and we focus on the aggregation of forecasts of probability distributions in Section 4. Section 5 provides a summary and concluding comments.

## 2. The COVID-19 Mortality Dataset

In this section, we introduce the dataset that is the focus of our study. We summarize the progression in mortality due to COVID-19 across the U.S. We then describe the forecasts in the dataset, and the criteria that we applied to select forecasts from this dataset for inclusion in our analysis.

The COVID-19 Forecast Hub is curated by a group led by Nicholas Reich (COVID-19 Forecast Hub 2020). The Hub provides open access to weekly observations and forecasts for the cumulative total of reported COVID-19 deaths, as well as observations and forecasts for the total deaths each week (incident deaths). These data are available for the U.S. both at the national and state levels. The forecasts are submitted by multiple forecasting teams from academia and industry.

### 2.1. Reported COVID-19 Mortality

The actual number of deaths from COVID-19 will be under-reported due to various factors including reporting delays and the lack of testing. There will also be indirect deaths as hospitals, overwhelmed by COVID-19, have had to delay diagnosis and treatment for other conditions, such as cancer. Therefore, the impact of COVID-19 will be judged ultimately in terms of the excess mortality, that is, the number of deaths above the number that would have been expected (Weinberger et al 2020). The actual cumulative deaths described in this paper are obtained from the COVID-19 Forecast Hub. These data are from the daily reports issued by the Centre for Systems Science and Engineering at Johns Hopkins University. Based on these data, the first reported death due to COVID-19 in the U.S. was in the State of Washington, in the week ending 29 February 2020. At the time of writing, 16 August 2020, the total number of COVID-19 deaths in the U.S. had risen to 169,481. Figure 1 shows the number of deaths across the U.S. on this date. Figure 2 shows the rise in the cumulative total number of COVID-19 deaths in the five states with highest cumulative total on this date.

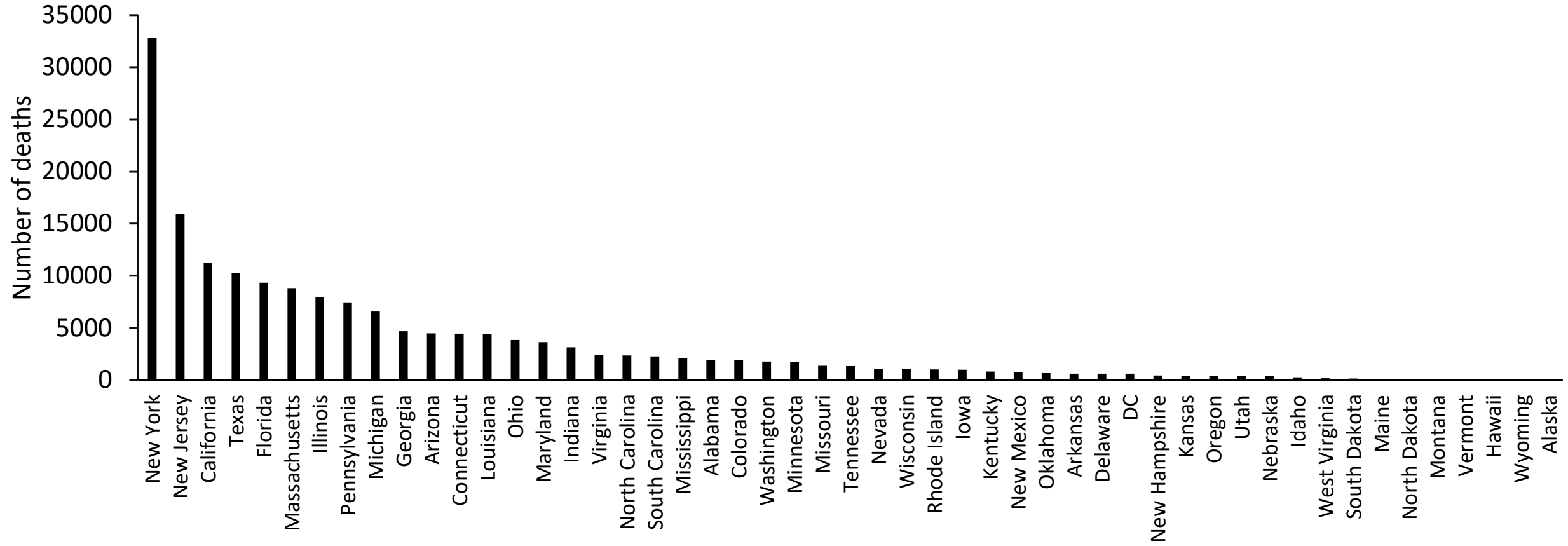

Figure 1: Number of reported COVID-19 deaths in the U.S. up to 16 August 2020.

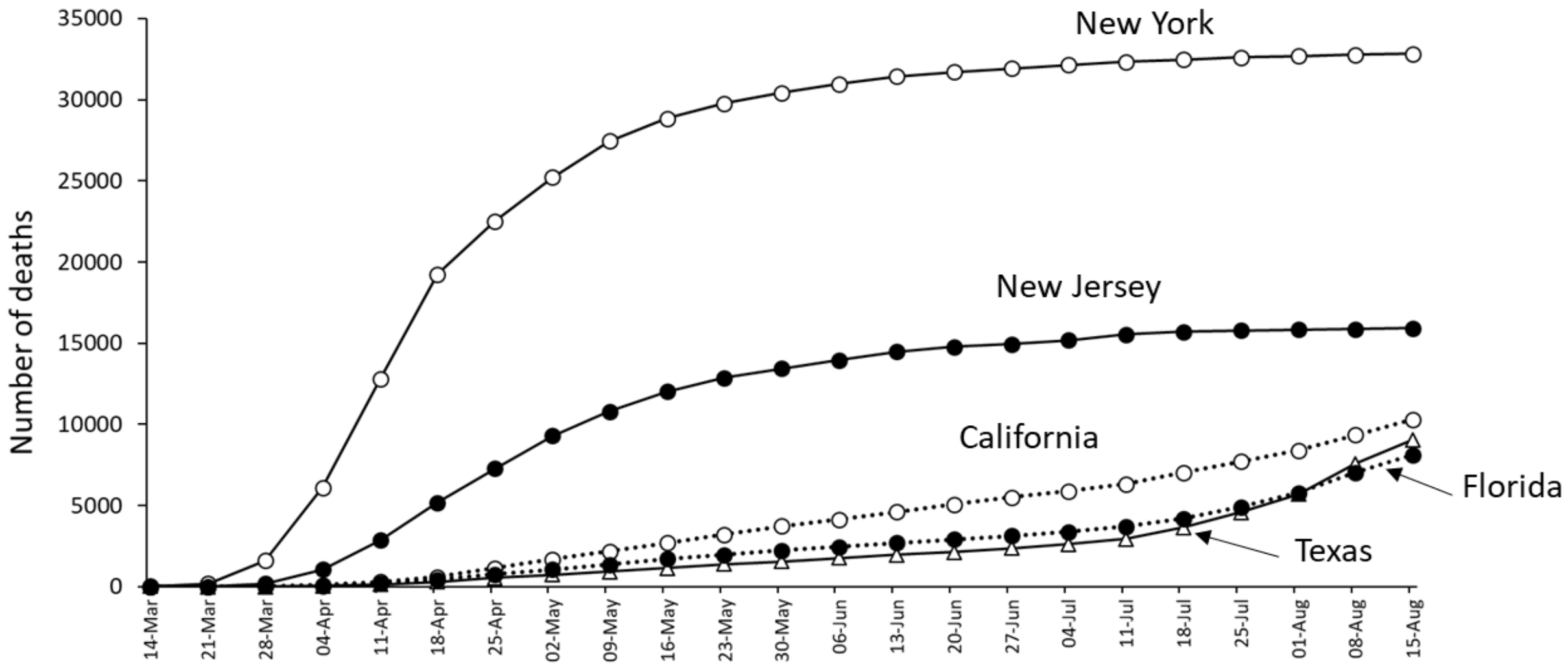

Figure 2: Rise in COVID-19 deaths in the five U.S. states with highest cumulative total up to 16 August 2020.

### 2.2. Forecasts of COVID-19 Mortality

The curators of the COVID-19 Forecast Hub ask forecasting teams to submit forecasts for one-week periods ending at midnight on Saturday evenings. The weeks are numbered starting with Week 0 defined as the week ending on Saturday 21 December 2019. At the end of each week, the number of incident and cumulative deaths is published, and with that week as forecast origin, the teams submit forecasts for 1 to 4 weeks ahead. For each of these lead times, and for incident and cumulative deaths, the teams provide a forecast of the probability distribution, which we refer to as the *distributional forecast*. It is provided in the form of forecasts of the quantiles corresponding to the following 23 probability levels: 1%, 2.5%, 5%, 10%, 15%, 20%, 25%, 30%, 35%, 40%, 45%, 50%, 55%, 60%, 65%, 70%, 75%, 80%, 85%, 90%, 95%, 97.5% and 99%. In addition, each team provides a point forecast of the central tendency of the distribution, which often coincides with their forecast of the median. The COVID-19 Forecast Hub provides data visualizations for the incident and cumulative numbers of deaths each week, with interactive plots so that the forecasts of different teams may be compared (see https://viz.covid19forecasthub.org/). Visualizations of the forecasts are also available from the website of the Centers for Disease Control and Prevention, and Nate Silver's FiveThirtyEight website.

As weekly cumulative and incident deaths are related, for simplicity, we focus only on cumulative deaths in our analysis. Although the COVID-19 Forecast Hub provides forecasts for all U.S. states and territories, we followed the convention adopted for the Hub's visualization by considering only the 50 states and the District of Columbia. For conciseness, we refer to these as 51 states. Given that we are also considering the national total, our dataset consists of 52 time series and associated forecasts.

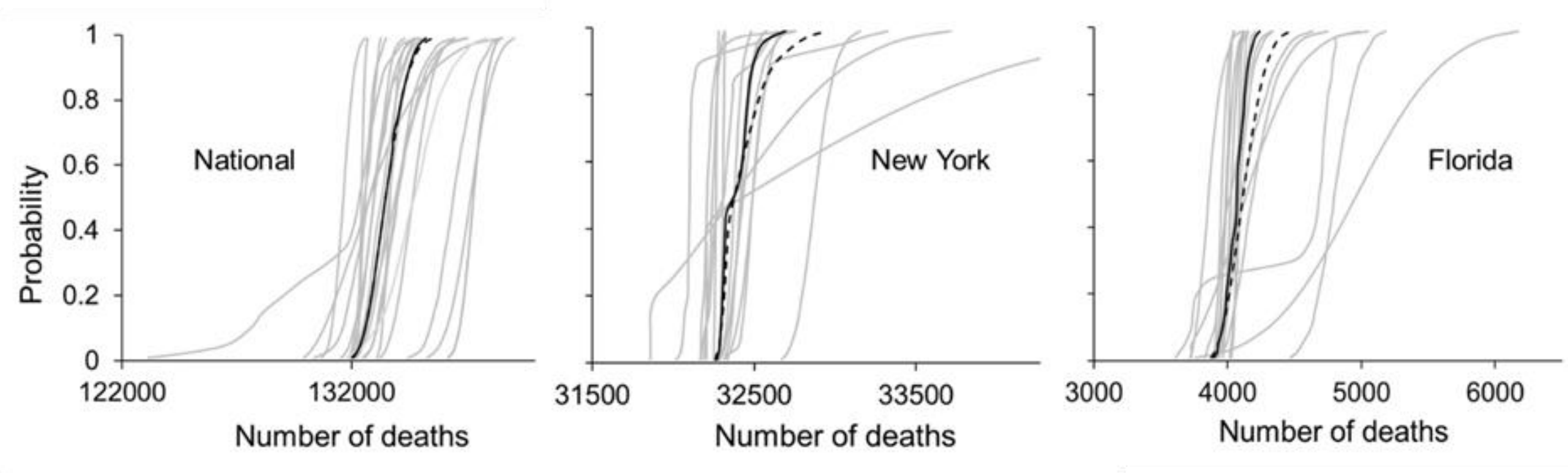

Figure 3: 1 week-ahead distributional forecasts produced with Week 28 as forecast origin.

In Figure 3, we give examples of distributional forecasts for the national level of cumulative mortality, and for the states of New York and Florida. In each plot, the distribution function shown as the dashed line is an aggregate forecast, proposed and made available by the curators of the Hub. They refer to it as the ensemble forecast, and, at the time of writing, it is constructed as the simple average of forecasts provided by the individual participating teams. The median distributional forecast is highlighted in black in each plot. Figure 3 shows that there is considerable variation among the distributional forecasts in terms of their location, spread and shape. The simple averaging provided by the ensemble forecast sits at the heart of each set of distributional forecasts. However, it is interesting to note that, at least for New York and Florida, there are a number of outlying distributional forecasts, which motivates consideration of alternative aggregation methods, based on robust estimation, such as the median and trimming. We discuss this further in Sections 3 and 4.

The COVID-19 Forecast Hub provides information regarding the methods used by the various forecasting teams and licensing conditions for the use of each team's data. The supplementary material to this paper summarizes this information. Approximately half of the teams use compartmental models. These involve the estimation of the rates at which the population, or sectors of the population, transfer between the states of being susceptible, exposed, infected and recovered/removed. Hence, they are widely referred to as SEIR or SIR models. The other forecasting teams used a variety of approaches, including agent-based simulation, statistical models, and deep learning. The use of data-driven machine learning methods is consistent with the increasing use of such methods in healthcare (see Guha and Kumar 2018). Surowiecki (2004) describes conditions under which wisdom can best be extracted from a crowd. These include independent contributors, diversity of opinions, and a trustworthy central convenor to collate the information provided. Forecasts from the Hub satisfy these conditions.

### 2.3. Inclusion Criteria for the Forecasts

In our analysis, we considered forecasts made with forecast origin as Week 18 and later. Week 18 seemed a reasonable starting point for our analysis because levels of mortality were relatively low prior to this, and it was the first forecast origin for which the ensemble forecast was included in the Hub's visualization. From Week 20, the curators of the Hub have produced files listing the forecasts that they did not include in their ensemble, based on several data screening checks. We omitted these forecasts from our analysis, and also followed the curators by treating as ineligible any submission that did not provide forecasts for all 23 quantiles and all four lead times. Unless this criterion was not met, we included forecasts not recorded as being assessed for eligibility because we felt we had no clear justification for omitting them. We considered data screening checks for these forecasts, but concluded that setting our own thresholds for inclusion would be arbitrary. Consequently, we applied aggregation methods to a dataset of forecasts that included 30% more forecasts than were included in the ensemble aggregation method.

Figure 4 shows the total number of forecasting teams included in our study for each forecast origin from Week 18 up to Week 33, which is the most recent origin at the time of writing. For each week, the figure also shows the split between the number of teams that used compartmental models and the number using alternatives. Note that the number of teams shown for each week in Figure 4 is an upper bound for the number available for aggregation for any one time series. This is because some teams either did not provide forecasts or did not provide eligible forecasts for all time series.

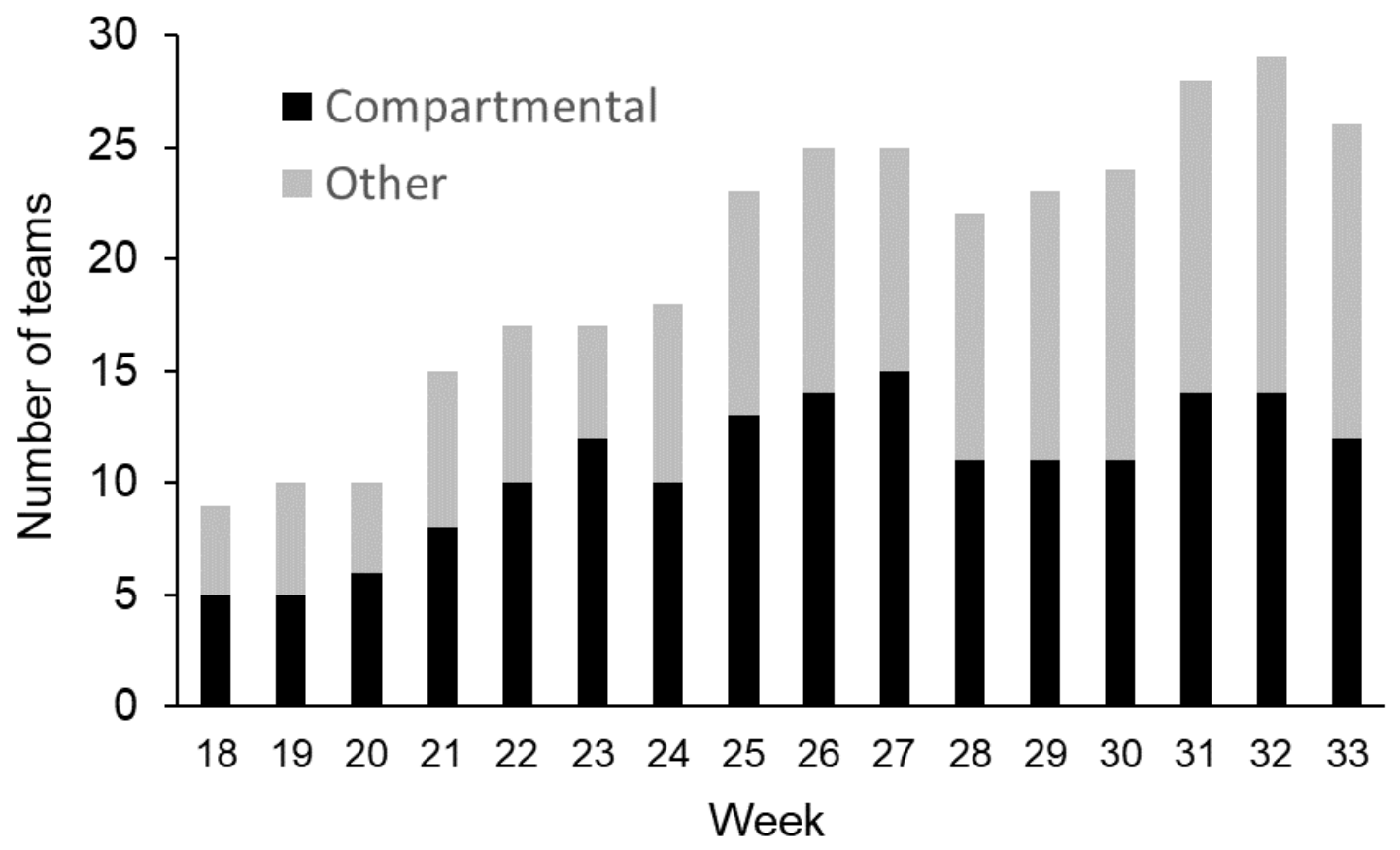

Figure 4: Number of forecasting teams included in our study for each forecast origin.
The stacked bars indicate the split between teams using compartmental models, and alternatives.

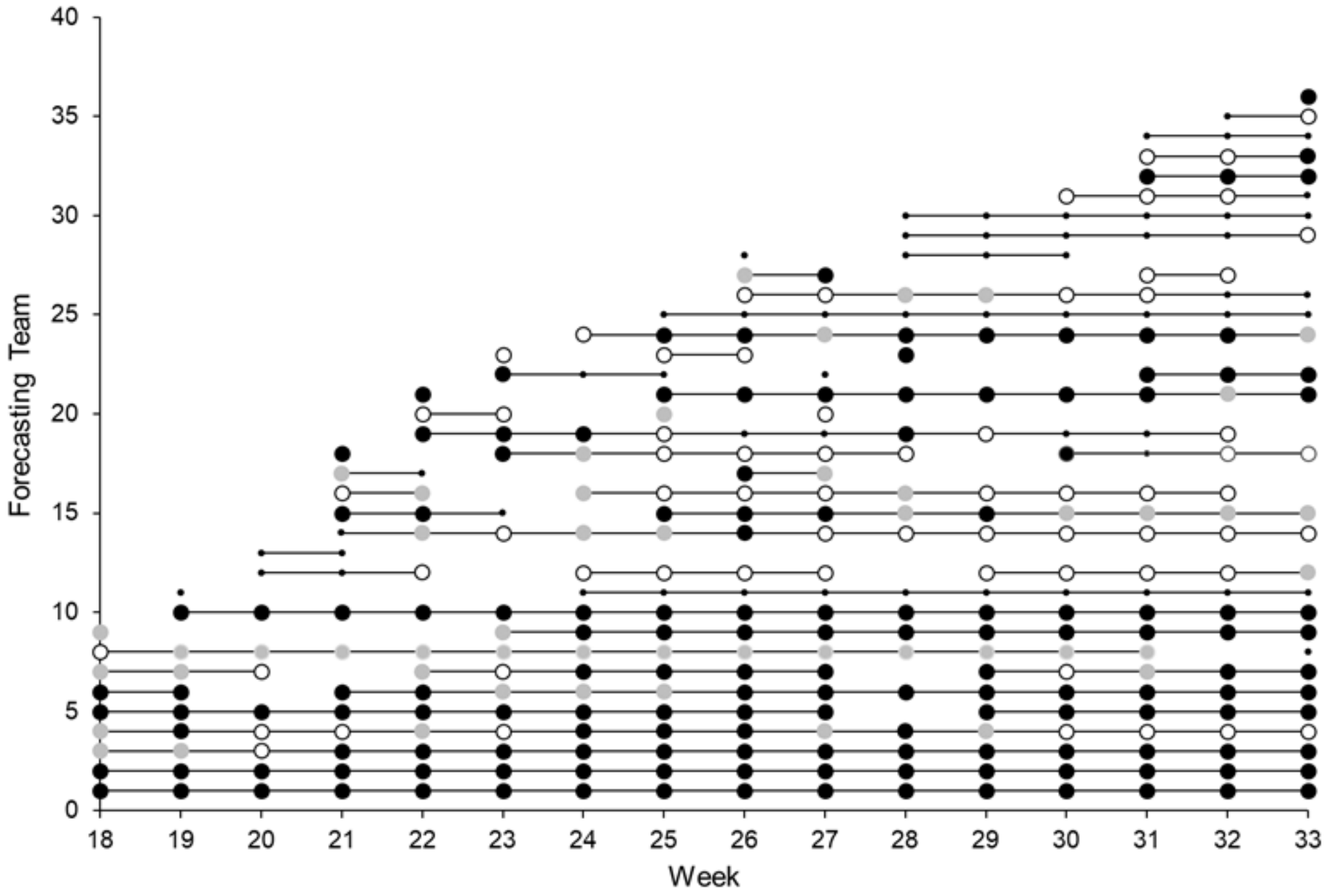

Figure 5: Timeline showing whether forecasts from each team were included in our study for each forecast origin.
The circles indicate the number of the 52 series for which forecasts were available and eligible. Black, grey, white and small black circles indicate: all 52 time series, either 51 or 50, between 49 and 26, and 25 or fewer, respectively.

For each of the 36 forecasting teams that submitted forecasts, Figure 5 shows, for each week, whether we were able to include that team in the aggregated forecasts for at least one series. A break in the horizontal line for any team indicates that, from the first week when the team submitted forecasts, it was not the case that forecasts from that team were available and eligible in all the following weeks. The circles in Figure 5 give an indication of the extent to which each forecasting team featured in our study. The figure shows that, even when a record of past historical accuracy becomes available for all teams, accuracy will not be available for the same past periods and same time series. This implies that weighted aggregation methods, based on the historical accuracy of each method, will not be straightforward to construct. Hence, our interest in aggregation methods that do not require a record of past accuracy for each team.

It would be interesting to compare the accuracy of the forecasts produced by the different teams. However, as Figure 5 shows that, for the full period that we use to compare the accuracy of the aggregation methods, we have forecasts for all 52 time series from only two of the participating teams. The results of these two teams did not compete with the results of the better aggregation methods.

## 3. Aggregating Interval Forecasts of COVID-19 Mortality

In this section, we describe the interval forecast aggregation methods that have been proposed in the literature, and then provide a summary list of the methods that we implemented. We then describe the measures that we use to evaluate interval forecasts, and finally use these to compare aggregation methods.

### 3.1. An Overview of Aggregation Methods for Interval Forecasts

We now briefly review methods for aggregating interval forecasts that are relevant to applications, such as ours, where there is a sizeable group of individual forecasters and a record of past accuracy is not available. The methods that have been proposed consider each bound of the interval separately. An obvious simplistic approach is to use the simple average of the forecasts. In the vast literature on combining point forecasts, it is well established that the simple average can be very competitive in a variety of applications. Interestingly, this is true regardless of whether a record of historical accuracy is available to enable far more sophisticated combining methods to be fitted (Larrick and Soll 2006). An advantage of the simple average is its simplicity, and robustness to changes over time in the relative performance of the individual methods. However, consideration of robustness prompts the use of aggregation methods that are robust to outliers, with the obvious candidate being the median. The simple average and the median are both considered by Park and Budescu (2015) and Gaba et al. (2017), who also propose several novel aggregation methods.

Extending the idea of robustness to outliers, Park and Budescu (2015) propose that, for each bound, a chosen percentage of the highest and lowest forecasts are discarded, followed by averaging of the rest. We refer to this as *symmetric trimming*. The median is an extreme version of symmetric trimming, where all but one forecast is trimmed.

Rather than having robustness as motivation, Gaba et al. (2017) use trimming to address the situation where the individual forecasters tend to be either under- or overconfident. We refer to their methods as *asymmetric trimming*. Their *exterior trimming* involves removing a percentage of the highest-valued upper bounds and lowest-valued lower bounds, with the aggregate computed by averaging the remaining upper and lower bounds, respectively. This approach is suitable when the forecasters have tended to be underconfident, with interval forecasts that are generally too wide. Gaba et al. (2017) also suggest *interior trimming*, which involves removing a percentage of the lowest-valued upper bounds and highest-valued lower bounds, followed by averaging. This approach to aggregation is suitable when the forecasters are overconfident. They also propose an extreme version of interior trimming, which involves discarding all but the highest upper bound and lowest lower bound. They refer to this as the *envelope* method. In addition, Gaba et al. (2017) describe a heuristic approach that views the bounds of each forecaster as having been produced based on a normal distribution.

The empirical studies of Park and Budescu (2015) involved 80% and 90% interval forecasts produced judgmentally by volunteers in experiments, where the interval forecasts related to general knowledge questions and estimates of financial and economic quantities. They found that the simple average and median were outperformed by trimming, and that symmetric was preferable to asymmetric trimming. Gaba et al. (2017) considered 90% interval forecasts produced for financial quantities by employees at a financial brokerage firm, and for macroeconomic variables by participants in the Federal Reserve Bank of Philadelphia's Survey of Professional Forecasters. These forecasts are produced using a variety of methods, including statistical models, expert judgment, and a mixture of the two. They find that exterior trimming performs very well for some of their data, but that the ranking of aggregation methods is dependent on the characteristics of the individual forecasts, such as under- or overconfidence. In a recent empirical paper, Grushka-Cockayne and Jose (2020) applied aggregation methods to 95% interval forecasts produced by statistical time series methods for the 100,000 series from the M4-Competition. Overall, the best results were achieved with median aggregation and interior asymmetric trimming.

### 3.2. Aggregation Methods Implemented in this Study

For each mortality series, forecast origin and lead time, we applied the following methods:

***Ensemble***: This is the simple average produced by the COVID-19 Forecast Hub. As discussed in Section 2, this is computed from a subset of the forecasts used in the other aggregation methods that we consider.

***Simple average***: For each bound, we computed the simple average of forecasts of this bound. We used our full set of forecasts for this and the other aggregation methods described below.

***Median***: For each bound, we computed the median of forecasts of this bound.

***Symmetric trimming*** $\boldsymbol{\beta}$: For each bound, we averaged the forecasts remaining after the removal of the $N$ lowest-valued and $N$ highest-valued forecasts, where $N$ is the largest integer less than or equal to the product of $\beta/2$ and the total number of forecasts. For all trimming methods in this study, we used $\beta$=10%, 20%, 30%, 40%, 50%, 60%, 70% 80% and 90%.

***Asymmetric exterior trimming*** $\boldsymbol{\beta}$: We first removed the $N$ lowest-valued lower bound forecasts, as well as the $N$ highest-valued upper bound forecasts, where $N$ is the largest integer less than or equal to the product of $\beta$ and the number of forecasts. For each bound, we averaged the remaining forecasts. This sometimes delivered a lower bound that was above the upper bound. In this situation, we replaced the two bounds by their mean. Although the resultant interval has zero width, it is preferable to having the upper bound above the lower.

***Asymmetric interior trimming*** $\boldsymbol{\beta}$: We removed the $N$ highest-valued lower bound forecasts, as well as the $N$ lowest-valued upper bounds, where $N$ is defined as for asymmetric exterior trimming. For each bound, we averaged the remaining forecasts.

***Envelope***: The interval is constructed using the lowest-valued lower bound forecast and highest-valued upper bound forecast.

### 3.3. Evaluation Measures for Interval Forecasts

In this paper, our interest is in central (1-$\alpha$) intervals. These are bounded by the $\alpha/2$ and (1-$\alpha/2$) quantiles. We consider $\alpha$=5% and 50%, which correspond to 95% and 50% intervals, respectively. Our choice of these intervals reflects those presented in the visualization of the COVID-19 Forecast Hub.

The accuracy of a set of interval forecasts can be assessed by evaluating the forecasts of the quantiles bounding the interval. A simple measure of accuracy of forecasts of the $\theta$ quantile is to check the percentage of observations that fall below the forecast. We refer to this as the *hit percentage*. If this is equal to $\theta$, the forecast is said to be *calibrated*. More precisely, we should refer to this as *unconditional calibration*, with *conditional calibration* being the property that the conditional expectation of the hit percentage is equal to $\theta$ (Nolde and Ziegel 2017). Given the short length of the time series considered in our analysis, we assess only unconditional calibration.

In addition to calibration, a quantile forecast should be evaluated using a *score*. A score can be viewed as a measure of how closely the forecast varies over time with the actual quantile. The score is said to be *consistent* if it is minimised by the true quantile. The use of a consistent score ensures honest reporting by a forecaster (Gneiting and Raftery, 2007). For quantile forecasts, the most widely used consistent score is the quantile regression loss function (see Koenker and Machado 1999; Taylor 1999). We refer to it as the *quantile score*, and present it as follows:

$$S_\theta^q\left(q_t(\theta), y_t\right) = \left(\theta - I\left\{y_t \le q_t(\theta)\right\}\right)\left(y_t - q_t(\theta)\right) \tag{1}$$

where $y_t$ is the observation in period $t$, $q_t(\theta)$ is the $\theta$ quantile, and $I\{\cdot\}$ is the indicator function. When $\theta$=50%, the score reduces to the absolute error, showing that the widely used mean absolute error is an appropriate score for a point forecast defined as a prediction of the median. To summarize forecasting performance across a time series, the average of the score is computed. A consistent score for an interval forecast is produced by summing the quantile score for the quantiles bounding the interval (Gneiting and Raftery 2007). For an interval bounded by $q_t(\alpha/2)$ and $q_t(1-\alpha/2)$, if we sum the quantile scores and divide by $\alpha/2$, we get the following *interval score* (Winkler, 1972):

$$S_\alpha^{INT}(l_t, u_t, y_t) = (u_t - l_t) + \frac{2}{\alpha} I\{y_t \le l_t\}(l_t - y_t) + \frac{2}{\alpha} I\{y_t \ge u_t\}(y_t - u_t) \quad (2)$$

where $l_t$ is the interval's lower bound $q_t(\alpha/2)$ and $u_t$ is its upper bound $q_t(1\text{-}\alpha/2)$. The score has the intuitive interpretation that it rewards narrow intervals, with observations that fall outside the interval incurring a penalty, the magnitude of which depends on the value of $\alpha$ (Gneiting and Raftery 2007). In an application to influenza forecasting, Bracher et al. (2020) use this interpretation to seek insight into why the interval score for one forecasting model is lower than another.

A notable aspect of forecast evaluation for our application is that we have a relatively small number of time periods, which is of particular concern when evaluating extreme quantiles, such as the 2.5% and 97.5% quantiles. In view of this, and the fact that the relative performances of the methods were similar across the four lead times, we opted to average the hit percentage and interval score across the lead times. In terms of averaging these measures over the 52 mortality series, this is unproblematic for calibration, but as the level of mortality varies greatly across the series, it is inevitable that averaging will lead to the interval score being dominated by its value for the high mortality series. In view of this, we report results for the following three groupings of the 52 series: the 22 states with cumulative mortality less than 1,000 at the end of the final week of our dataset; the 25 states with more than 1,000 and less than 10,000 at the end of the final week; and the series with at least 10,000 at the end of the final week, which corresponded to the four hardest hit states and the national U.S. series. We refer to these groupings as the *low*, *medium* and *high* mortality series.

### 3.4. Interval Forecasting Results

For the three groupings, the mean of the interval score for 95% interval forecasts is presented in the first three columns of values in Table 1, and their ranks are provided in the next three columns. The unit of the score is deaths, and lower values of the score reflects greater accuracy. The final three columns present the skill score, which measures the percentage by which the score for a given method is better than a benchmark method, which we chose to be the simple average. For conciseness, we present the results for trimming with $\beta$=20%, 40%, 60% and 80%, as we feel this provides an adequate summary. For the low and medium mortality series, comfortably the best results were obtained using the median aggregation method and symmetric trimming with $\beta$=40%, 60% and 80%. For these series, the ensemble, the simple average and asymmetric trimming performed quite poorly. However, for the high mortality series, the best results were produced by the simple average, with symmetric trimming and asymmetric interior trimming also performing well. For the high mortality series, a curious finding is that the simple average of our set of forecasts outperforms the ensemble forecast, which is a simple average of a narrower set of forecasts. Including the additional information seems to have benefitted our simple average aggregation. For all series, the envelope aggregation method produced poor results.

Table 1: Mean interval score for 95% interval forecasts.

| | Interval Score | | | Rank of Score | | | Skill Score (%) | | |
|---|---|---|---|---|---|---|---|---|---|
| | Low | Medium | High | Low | Medium | High | Low | Medium | High |
| Ensemble | 184 | 950 | 13275 | 9 | 8 | 8 | -4.9 | -0.2 | -11.8 |
| Simple average | 175 | 948 | 11872 | 7 | 7 | 3 | 0.0 | 0.0 | 0.0 |
| Median | 85 | 570 | 13598 | 1 | 4 | 10 | 51.6 | 39.9 | -14.5 |
| Sym trim 20% | 157 | 792 | 12846 | 6 | 6 | 6 | 10.6 | 16.4 | -8.2 |
| Sym trim 40% | 95 | 563 | 12833 | 4 | 3 | 5 | 45.8 | 40.6 | -8.1 |
| Sym trim 60% | 88 | 554 | 13255 | 3 | 1 | 7 | 49.5 | 41.5 | -11.6 |
| Sym trim 80% | 86 | 559 | 13486 | 2 | 2 | 9 | 50.9 | 41.0 | -13.6 |
| Asym ext trim 20% | 127 | 683 | 15905 | 5 | 5 | 12 | 27.6 | 27.9 | -34.0 |
| Asym ext trim 40% | 198 | 1232 | 21287 | 10 | 10 | 14 | -13.2 | -30.0 | -79.3 |
| Asym ext trim 60% | 353 | 2164 | 29900 | 13 | 13 | 15 | -101.5 | -128.3 | -151.8 |
| Asym ext trim 80% | 557 | 3613 | 40239 | 15 | 15 | 16 | -218.0 | -281.3 | -238.9 |
| Asym int trim 20% | 183 | 1056 | 11123 | 8 | 9 | 1 | -4.3 | -11.4 | 6.3 |
| Asym int trim 40% | 229 | 1329 | 11341 | 11 | 11 | 2 | -30.6 | -40.3 | 4.5 |
| Asym int trim 60% | 309 | 1755 | 12236 | 12 | 12 | 4 | -76.3 | -85.2 | -3.1 |
| Asym int trim 80% | 495 | 2753 | 14677 | 14 | 14 | 11 | -182.8 | -190.6 | -23.6 |
| Envelope | 843 | 4604 | 21072 | 16 | 16 | 13 | -381.5 | -385.8 | -77.5 |

Note: The unit of the score is deaths. Lower values of the score and higher values of the skill score are better.

Table 2 presents the mean of the interval score for 50% interval forecasts. The results are broadly consistent with those for the 95% interval. Symmetric trimming and the median method were the most accurate for the low and medium mortality series. For the high mortality series, the best results were achieved with the simple average.

Figures 6 to 8 summarize calibration by using Q-Q plots to report the hit percentages for the bounds of both the 50% and 95% interval forecasts. To ensure readability of the plots, we present the results for just the ensemble, simple average, median and two of the symmetric trimming methods, as these were competitive in terms of the interval score in Tables 1 and 2. In Figures 6 to 8, the dashed line indicates the ideal performance. Figures 6 and 7 show that, for each method applied to the low and medium mortality series, forecasts of the 2.5% and 97.5% quantile were very close to the ideal. In these figures, for the 25% and 75% quantiles, the best results were for symmetric trimming, and for the median method. Figure 8 shows that, for the high mortality series, all the methods produced forecasts of the 25%, 75% and 97.5% quantiles that were too low.

Table 2: Mean interval score for 50% interval forecasts.

| | Interval Score | | | Rank of Score | | | Skill Score (%) | | |
|---|---|---|---|---|---|---|---|---|---|
| | Low | Medium | High | Low | Medium | High | Low | Medium | High |
| Ensemble | 54 | 311 | 3695 | 8 | 6 | 9 | 1.3 | 2.5 | -7.4 |
| Simple average | 55 | 319 | 3439 | 9 | 8 | 1 | 0.0 | 0.0 | 0.0 |
| Median | 41 | 290 | 3550 | 2 | 2 | 4 | 25.4 | 9.1 | -3.2 |
| Sym trim 20% | 49 | 296 | 3535 | 5 | 5 | 3 | 9.5 | 7.0 | -2.8 |
| Sym trim 40% | 43 | 289 | 3535 | 4 | 1 | 2 | 21.9 | 9.5 | -2.8 |
| Sym trim 60% | 41 | 290 | 3555 | 3 | 3 | 5 | 25.1 | 9.0 | -3.4 |
| Sym trim 80% | 40 | 291 | 3558 | 1 | 4 | 6 | 26.0 | 8.8 | -3.5 |
| Asym ext trim 20% | 50 | 313 | 3573 | 6 | 7 | 7 | 8.4 | 1.9 | -3.9 |
| Asym ext trim 40% | 54 | 347 | 3836 | 7 | 9 | 10 | 1.8 | -8.9 | -11.6 |
| Asym ext trim 60% | 58 | 373 | 3993 | 11 | 11 | 11 | -5.7 | -17.1 | -16.1 |
| Asym ext trim 80% | 71 | 438 | 4283 | 13 | 12 | 13 | -29.3 | -37.5 | -24.5 |
| Asym int trim 20% | 56 | 355 | 3641 | 10 | 10 | 8 | -2.8 | -11.4 | -5.9 |
| Asym int trim 40% | 70 | 449 | 4142 | 12 | 13 | 12 | -28.0 | -40.8 | -20.5 |
| Asym int trim 60% | 90 | 580 | 4928 | 14 | 14 | 14 | -65.6 | -81.9 | -43.3 |
| Asym int trim 80% | 128 | 802 | 6275 | 15 | 15 | 15 | -135.5 | -151.5 | -82.5 |
| Envelope | 198 | 1212 | 8807 | 16 | 16 | 16 | -262.1 | -280.4 | -156.1 |

Note: The unit of the score is deaths. Lower values of the score and higher values of the skill score are better.

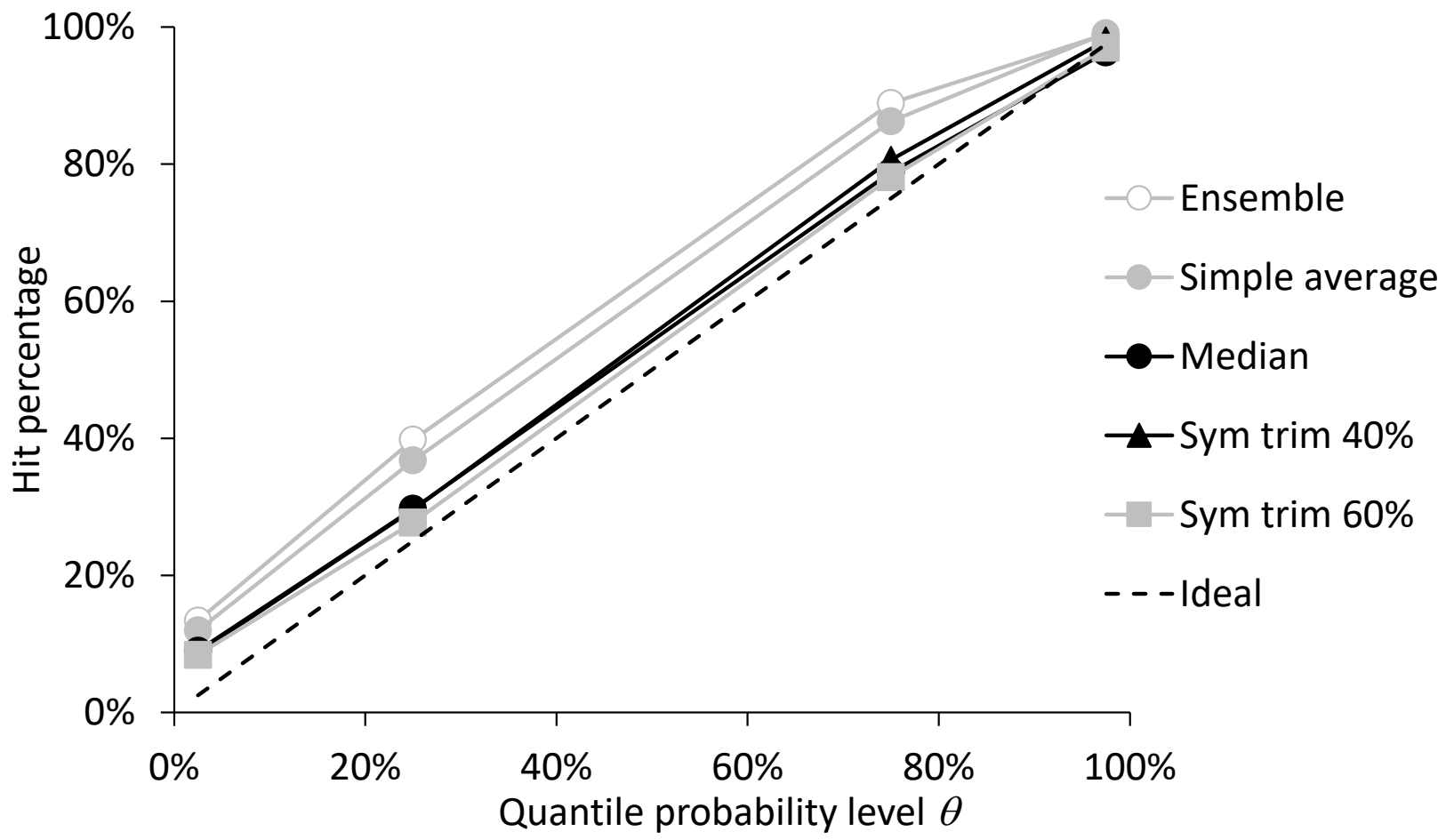

Figure 6: Calibration hit percentages for bounds on 50% and 95% intervals for the low mortality series.

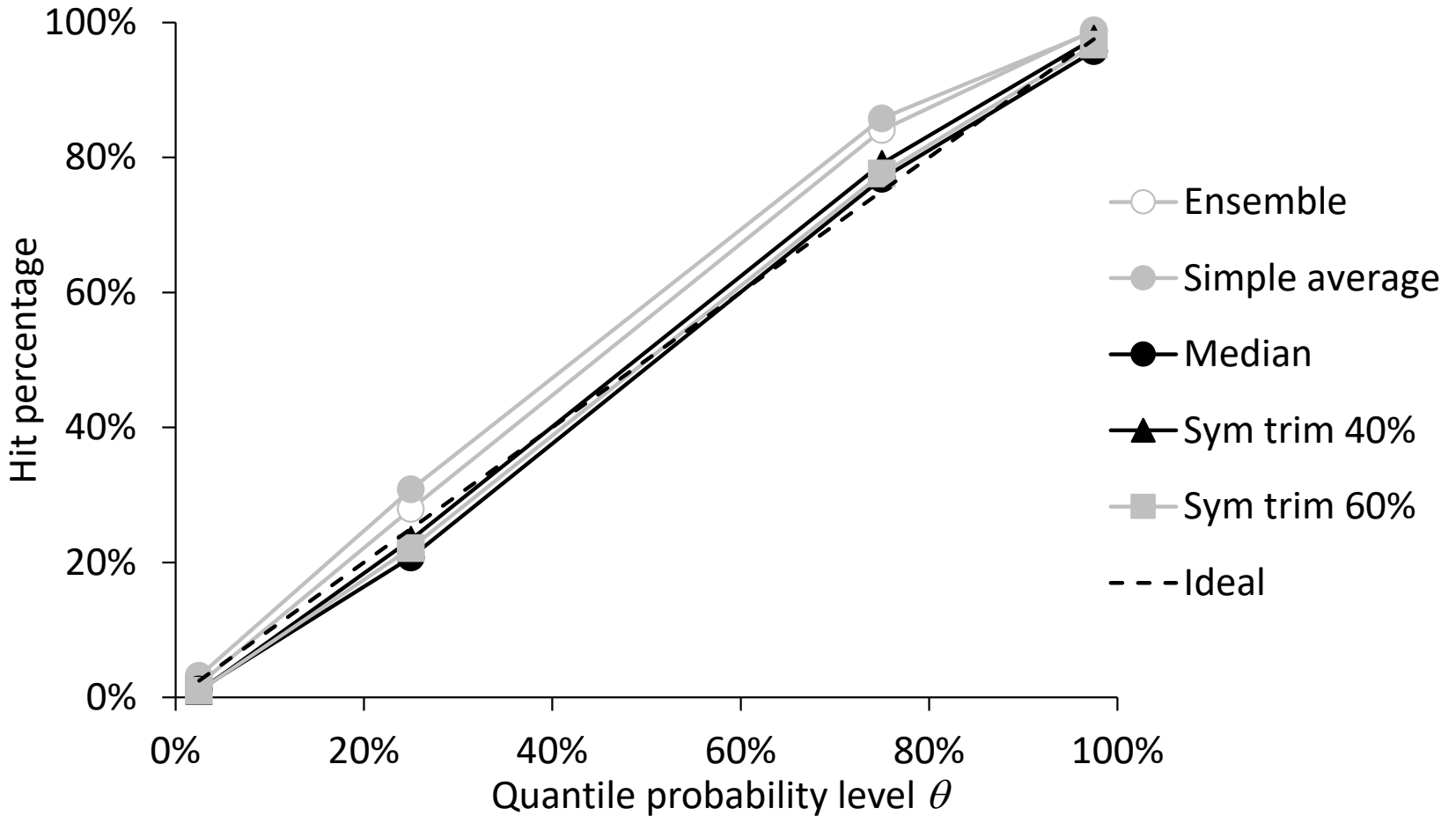

Figure 7: Calibration hit percentages for bounds on 50% and 95% intervals for the medium mortality series.

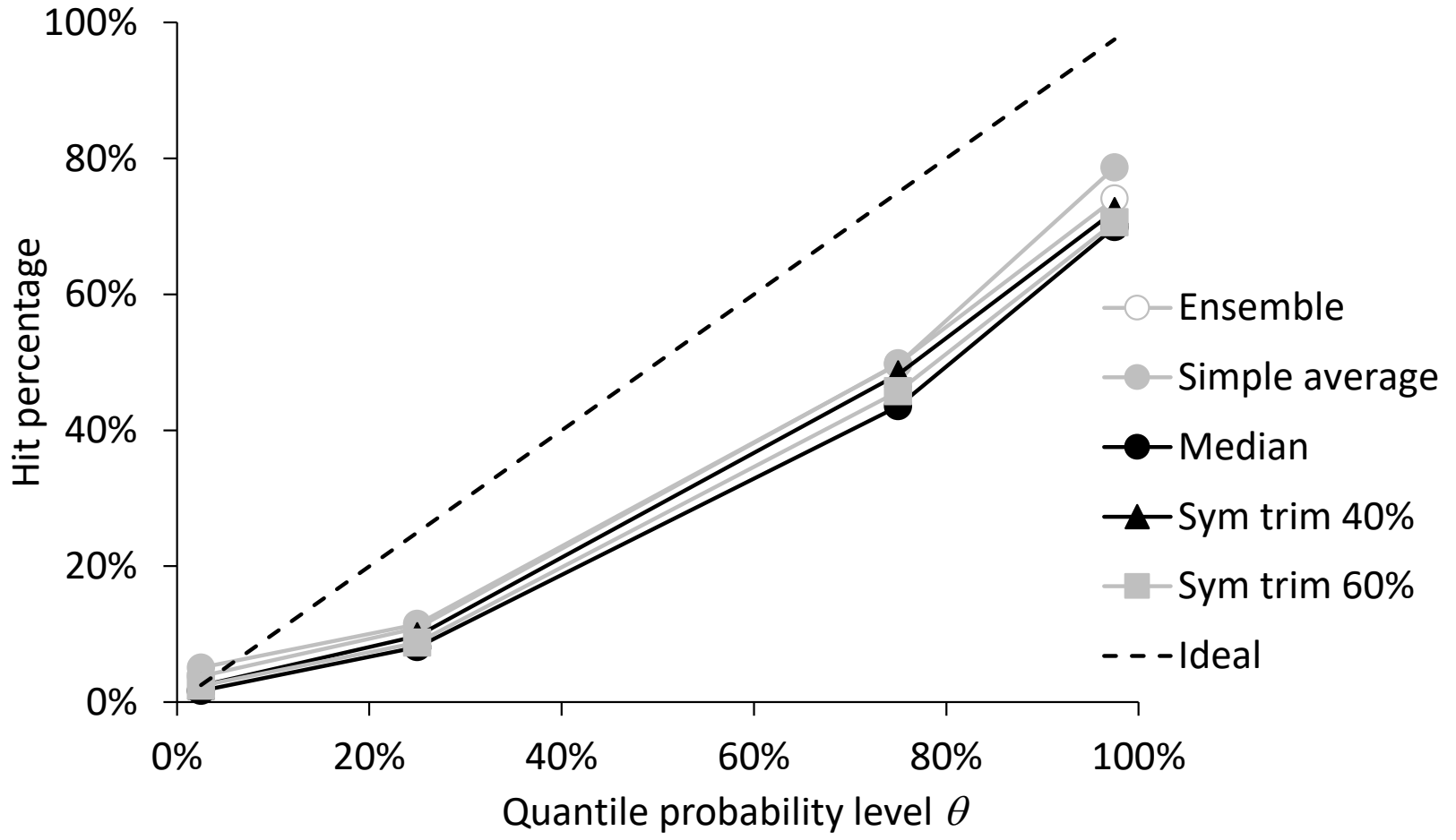

Figure 8: Calibration hit percentages for bounds on 50% and 95% intervals for the high mortality series.

## 4. Aggregating Distributional Forecasts of COVID-19 Mortality

In this section, we briefly review the distributional forecast aggregation methods that have been proposed in the literature, and then list the methods that we implemented. We then describe measures for evaluating distributional forecast accuracy, before using these to compare different aggregation methods.

### 4.1. An Overview of Aggregation Methods for Distributional Forecasts

The simple average is a well-established approach for aggregating distributional forecasts (see, for example, Stone 1961). It has typically taken the form of the linear opinion pool, which, for any chosen value of the random variable, is the average of the corresponding cumulative probabilities obtained from the distributional forecasts. However, this form of averaging has been criticized because, even when the quantiles of the individual distributions are calibrated, the linear opinion pool will not itself provide perfect calibration (Hora 2004, Ranjan and Gneiting 2010). It has been noted that, when there is diversity among the means of the individual forecasts, this will tend to lead to an exaggerated variance in the forecast of the linear opinion pool (see, for example, Dawid 1995). To address these problems, Lichtendahl et al. (2013) propose that, instead of averaging the cumulative probabilities, the distributional forecasts should be averaged by taking the mean of their corresponding quantile forecasts. In other words, for a chosen probability level $\theta$, they suggest that aggregation is performed by averaging forecasts of the quantile $q_t(\theta)$ provided by each individual distributional forecast. They show how this leads to more attractive theoretical properties than the linear opinion pool. For our application in this paper, it provides a more convenient approach to averaging because each forecasting team provides the distributional forecast in terms of quantile forecasts for the same 23 values of the probability $\theta$.

To provide robust aggregation, Hora et al. (2013) propose that the median is used. For any chosen value of the random variable, their approach finds the median of the corresponding values of the cumulative probability forecasts. In fact, they show that this delivers the same aggregate distributional forecast as an approach that uses the median of forecasts of the quantile $q_t(\theta)$ for a chosen probability level $\theta$.

As with interval forecast aggregation, trimming has also been proposed for distributional forecasts. Jose et al. (2014) propose *interior* and *exterior* trimming approaches, which involve trimming the innermost and outermost distributions, respectively, from a set of distributional forecasts. To enable the trimming, the distributional forecasts must essentially be ordered in some way, and for this, they propose two alternative approaches: the *CDF approach* (CA) and *mean approach* (MA). MA orders the distributional forecasts according to their means, and involves trimming entire distributional forecasts. After a proportion have been trimmed, the authors use a linear opinion pool to average the rest. CA orders the distributional forecasts separately for each of a set of values of the random variable. After the trimming is performed, the aggregate forecast is computed as the average of the cumulative probabilities given by the remaining distributional forecasts. Jose et al. (2014) note that CA could be adapted so that the trimming and averaging is performed on forecasts of the quantile $q_t(\theta)$ for any chosen value of the probability $\theta$, which would be more consistent with the advice of Lichtendahl et al. (2013) to average quantiles, rather than probabilities. For our application, this is a more convenient way to implement CA, as our distributional forecasts have each been provided in the form of a set of quantile forecasts. Following similar reasoning, we avoided the linear opinion pool with MA, so that following trimming, quantile forecasts are averaged.

For interval forecast aggregation, symmetric trimming is motivated by robustness, and asymmetric trimming enables the impact to be reduced of a tendency among the individual forecasters to be either under- or overconfident. It is worth noting that analogous asymmetric methods are not straightforward for distributional forecasts because of the need to ensure that the resulting distribution function is monotonically increasing. It is also interesting to note that, although the trimming methods proposed by Jose et al. (2014) are all symmetric, and hence their exterior trimming will enable the removal of outliers, their main motivation for trimming is to address under- or overconfidence among the individual forecasters. For example, in their empirical application to forecasts from the Federal Reserve Bank of Philadelphia's Survey of Professional Forecasters, they show that, in comparison with the linear opinion pool, exterior trimming enables the impact to be reduced of underconfident forecasts of inflation, while interior trimming allows the aggregate distributional forecast to reduce the impact of overconfident forecasts of growth. Grushka-Cockayne et al. (2017) investigate the theoretical properties of CA with exterior trimming, and compare it with the linear opinion pool. They show that exterior trimming can overcome the tendency for the linear opinion pool to produce an underconfident forecast when the individual distributional forecasts have diverse means. They demonstrate this with an ensemble of distributional forecasts from a quantile regression forest. They show that the tendency for this machine learning method to overfit leads to diverse means among the ensemble, which can be addressed by CA with exterior trimming.

### 4.2. Aggregation Methods Implemented in this Study

For each mortality series, forecast origin and lead time, we implemented the following methods:

***Ensemble***: This is the simple average produced by the COVID-19 Forecast Hub. For each of the 23 quantiles, the average is computed of the corresponding quantile forecasts obtained from the individual distributional forecasts. As we explained in Section 2, the ensemble used a subset of the forecasts included in all other aggregation methods that we considered.

***Simple average***: For each of the 23 quantiles, we averaged the corresponding quantile forecasts. We used our full set of forecasts for this and the other aggregation methods described below.

***Median***: For each of the 23 quantiles, we found the median of the quantile forecasts.

***CA exterior trimming $\beta$***: For each of the 23 quantiles, we computed the aggregate as the average of the quantile forecasts remaining after we had removed the $N$ lowest-valued and $N$ highest-valued quantile forecasts, where $N$ is the largest integer less than or equal to the product of $\beta/2$ and the total number of forecasts. For each bound, we averaged the remaining forecasts.

***CA interior trimming $\beta$***: With this method, for each of the 23 quantiles, the innermost quantile forecasts were trimmed. The aggregate was computed as the average of the quantile forecasts that were either among the $N$ lowest-valued or $N$ highest-valued quantile forecasts, where $N$ is the largest integer less than or equal to the product of $(1-\beta)/2$ and the total number of forecasts.

***MA exterior trimming $\beta$***: This method involved trimming entire distributional forecasts. The trimming was based on the mean of each distributional forecast, which we estimated using the average of the 23 quantile forecasts. The aggregate forecast was computed by averaging the distributional forecasts that remain after the removal of the $N$ distributional forecasts with lowest-valued mean and the $N$ distributional forecasts with highest-valued mean, where $N$ is the largest integer less than or equal to the product of $\beta/2$ and the total number of forecasts.

***MA interior trimming $\beta$***: This was similar to MA exterior trimming, except the innermost distributional forecasts were trimmed. The aggregate was computed as the average of the distributional forecasts that were among the $N$ distributional forecasts with lowest-valued mean and the $N$ distributional forecasts with highest-valued mean, where $N$ is the largest integer less than or equal to the product of $(1-\beta)/2$ and the total number of forecasts.

### 4.3. Evaluation Measures for Distributional Forecasts

Gneiting et al. (2007) describe how the aim of distributional forecasting is to maximise *sharpness* subject to *calibration*. Sharpness concerns the concentration of the distributional forecast, and calibration assesses its statistical consistency with the data. Randomly sampled values from a calibrated distributional forecast are indistinguishable from the observations (Gneiting and Katzfuss, 2014). To evaluate calibration for distributional forecasts in our study, we computed the hit percentage for each of the 23 quantiles.

A score summarises calibration and sharpness, and is said to be *proper* if it is minimized by the true distribution. As with consistent scoring functions for quantiles, proper distributional scores are recommended to ensure forecasters report honest predictions (Gneiting and Raftery 2007). A widely used proper score for distributions of continuous random variables is the continuous ranked probability score (CRPS) (see Gneiting and Raftery 2007). It can be viewed in several different ways, including the integral of the quantile score of expression (1), with respect to the probability level $\theta$. For our application, where we have quantile forecasts for just 23 different values of $\theta$, we approximate the CRPS using expression (3), which is the sum of the quantile scores for the 23 quantile forecasts:

$$S_{\theta_i}^{DIST}\left(q_t(\theta_i), y_t\right) = \sum_{i=1}^{23}\left(\theta_i - I\left\{y_t \le q_t(\theta_i)\right\}\right)\left(y_t - q_t(\theta_i)\right) \qquad (3)$$

This is a proper score, and this can be seen by viewing it as a quantile-weighted version of the CRPS (see Gneiting and Ranjan 2011). For simplicity, in this paper, we refer to expression (3) as the CRPS. We note that Bracher et al. (2020) present it as a form of weighted interval score, because it can be written as a weighted sum of the interval score of expression (2) and the quantile score of expression (1) for the median.

Although our main interest in this paper is probabilistic forecasting, we also consider point predictions of the median, as this conveys the accuracy of the centre of location of the distributional forecasts, and of course such point forecasts are often the main focus of attention. We evaluate these point forecasts using the mean absolute error (MAE).

**4.4. Distributional Forecasting Results**

Table 3 presents the mean of the CRPS for the three groupings of series, along with ranks and skill scores. The unit of the CRPS is the number of deaths, and lower values of the score reflect greater accuracy. For the low and medium mortality series, the best results were produced by median aggregation and the exterior trimming aggregation methods, regardless of whether the trimming was performed within the CA or MA methods. In view of our comments in Section 4.1, the usefulness of exterior trimming suggests either outliers or that the forecasting teams were generally underconfident. For the high mortality series, the best results were obtained using the simple average and the trimming methods with low $\beta$. As with the interval forecasting, it is notable that, for the high mortality series, the simple average of our set of forecasts outperformed the Hub's ensemble, which is a simple average of a more selective set of forecasts.

In Table 4, we evaluate point forecast accuracy for the median using the MAE. The unit of the MAE is the number of deaths. The relative performances of the methods in Table 4 are similar to those in Table 3 for the CRPS. This is quite common when there is sizeable variation over time in the location of the probability distribution, because inaccuracy in the prediction of the distribution's location will affect the accuracy of all the quantiles, and hence the whole distribution.

Table 3: Mean of the CRPS for the distributional forecasts.

| | CRPS | | | Rank of CRPS | | | Skill Score (%) | | |
|---|---|---|---|---|---|---|---|---|---|
| | Low | Medium | High | Low | Medium | High | Low | Medium | High |
| Ensemble | 129 | 737 | 8755 | 10 | 10 | 15 | 0.2 | 2.7 | -9.6 |
| Simple average | 129 | 758 | 7988 | 11 | 11 | 1 | 0.0 | 0.0 | 0.0 |
| Median | 94 | 657 | 8531 | 2 | 1 | 13 | 27.2 | 13.3 | -6.8 |
| CA ext trim 20% | 117 | 693 | 8309 | 8 | 8 | 6 | 9.3 | 8.5 | -4.0 |
| CA ext trim 40% | 98 | 660 | 8365 | 6 | 4 | 8 | 23.7 | 12.9 | -4.7 |
| CA ext trim 60% | 94 | 659 | 8456 | 3 | 3 | 10 | 26.9 | 13.0 | -5.9 |
| CA ext trim 80% | 92 | 658 | 8492 | 1 | 2 | 12 | 28.2 | 13.1 | -6.3 |
| CA int trim 20% | 150 | 833 | 8138 | 13 | 13 | 2 | -16.5 | -10.0 | -1.9 |
| CA int trim 40% | 181 | 956 | 8368 | 15 | 14 | 9 | -40.3 | -26.2 | -4.8 |
| CA int trim 60% | 246 | 1302 | 9452 | 17 | 17 | 16 | -91.3 | -71.9 | -18.3 |
| CA int trim 80% | 332 | 1973 | 12323 | 19 | 18 | 18 | -158.0 | -160.5 | -54.3 |
| MA ext trim 20% | 117 | 694 | 8293 | 9 | 9 | 5 | 8.8 | 8.3 | -3.8 |
| MA ext trim 40% | 102 | 668 | 8292 | 7 | 6 | 4 | 20.8 | 11.9 | -3.8 |
| MA ext trim 60% | 97 | 662 | 8332 | 4 | 5 | 7 | 24.5 | 12.6 | -4.3 |
| MA ext trim 80% | 98 | 676 | 8461 | 5 | 7 | 11 | 24.0 | 10.7 | -5.9 |
| MA int trim 20% | 148 | 833 | 8207 | 12 | 12 | 3 | -15.0 | -10.0 | -2.7 |
| MA int trim 40% | 177 | 958 | 8535 | 14 | 15 | 14 | -37.1 | -26.4 | -6.8 |
| MA int trim 60% | 233 | 1293 | 9880 | 16 | 16 | 17 | -81.0 | -70.7 | -23.7 |
| MA int trim 80% | 320 | 1985 | 13045 | 18 | 19 | 19 | -148.4 | -162.1 | -63.3 |

Note: The unit of the score is deaths. Lower values of the score and higher values of the skill score are better.

Table 4: MAE for the median point forecasts derived from the distributional forecasts.

| | MAE | | | Rank of MAE | | | Skill Score (%) | | |
|---|---|---|---|---|---|---|---|---|---|
| | Low | Medium | High | Low | Medium | High | Low | Medium | High |
| Ensemble | 16 | 94 | 1065 | 10 | 10 | 15 | 1.6 | 4.0 | -13.7 |
| Simple average | 16 | 98 | 937 | 11 | 11 | 1 | 0.0 | 0.0 | 0.0 |
| Median | 13 | 88 | 1016 | 2 | 1 | 12 | 21.3 | 9.7 | -8.5 |
| CA ext trim 20% | 15 | 90 | 973 | 8 | 7 | 3 | 8.9 | 7.7 | -3.9 |
| CA ext trim 40% | 13 | 89 | 988 | 6 | 4 | 7 | 18.0 | 8.8 | -5.4 |
| CA ext trim 60% | 13 | 89 | 1000 | 3 | 3 | 10 | 21.3 | 9.1 | -6.7 |
| CA ext trim 80% | 12 | 89 | 1004 | 1 | 2 | 11 | 23.0 | 9.3 | -7.2 |
| CA int trim 20% | 18 | 105 | 979 | 13 | 13 | 5 | -13.0 | -7.6 | -4.5 |
| CA int trim 40% | 21 | 117 | 1024 | 15 | 15 | 13 | -30.9 | -19.5 | -9.3 |
| CA int trim 60% | 27 | 148 | 1192 | 17 | 17 | 16 | -67.1 | -50.9 | -27.2 |
| CA int trim 80% | 36 | 217 | 1524 | 19 | 19 | 18 | -123.5 | -121.3 | -62.7 |
| MA ext trim 20% | 15 | 91 | 971 | 9 | 9 | 2 | 8.5 | 7.0 | -3.7 |
| MA ext trim 40% | 14 | 91 | 977 | 7 | 8 | 4 | 14.9 | 7.4 | -4.3 |
| MA ext trim 60% | 13 | 90 | 996 | 4 | 5 | 9 | 18.7 | 8.4 | -6.3 |
| MA ext trim 80% | 13 | 90 | 996 | 5 | 6 | 8 | 18.2 | 8.0 | -6.3 |
| MA int trim 20% | 18 | 105 | 984 | 12 | 12 | 6 | -11.3 | -7.5 | -5.1 |
| MA int trim 40% | 21 | 117 | 1034 | 14 | 14 | 14 | -28.2 | -19.3 | -10.4 |
| MA int trim 60% | 26 | 146 | 1232 | 16 | 16 | 17 | -58.4 | -49.2 | -31.5 |
| MA int trim 80% | 34 | 215 | 1564 | 18 | 18 | 19 | -112.2 | -119.5 | -66.9 |

Note: The unit of the score is deaths. Lower values of the score and higher values of the skill score are better. Bold indicates the best three methods in a column.

For the three groupings of series, Figures 9 to 11 present Q-Q plots to summarize the calibration hit percentages for each of the 23 quantiles. To ensure legibility of the figures, we include just the following five methods in each: the ensemble, the simple average, the median, and two of the more successful external trimming methods. Figure 9 shows that, for the low mortality series, on average, the quantiles of the distributional forecasts were too high, while for the high mortality series, Figure 11 shows that the quantile forecasts were generally too low. In Figures 9 and 10, we see that for the low and medium mortality series, the best calibration results correspond to median aggregation and exterior trimming. For the high mortality series, Figure 11 shows that the best calibration corresponds to the ensemble and simple average.

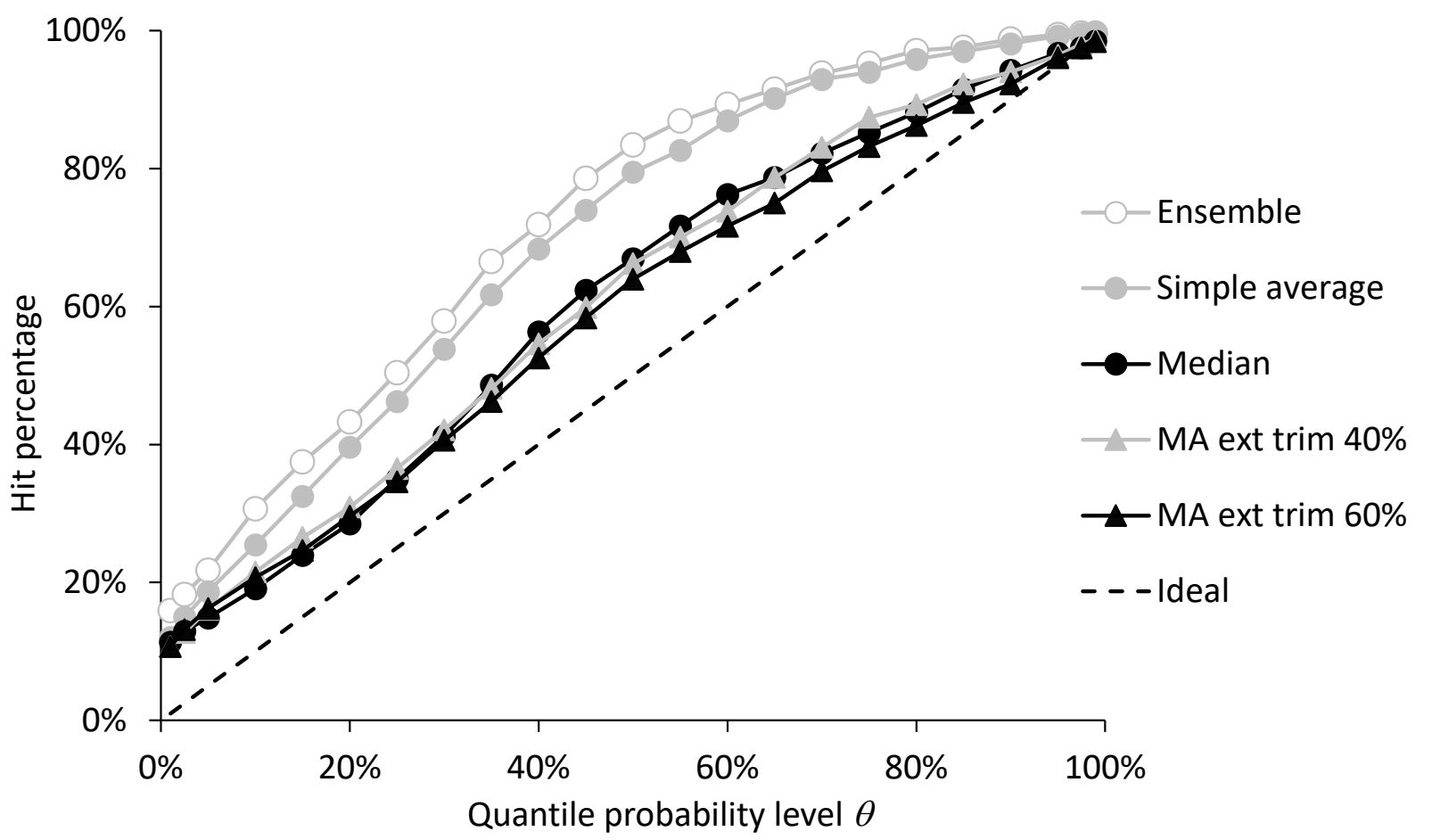

Figure 9: Calibration of distributional forecasts assessed using hit percentages for the 23 quantile probability levels $\theta$ for the low mortality series.

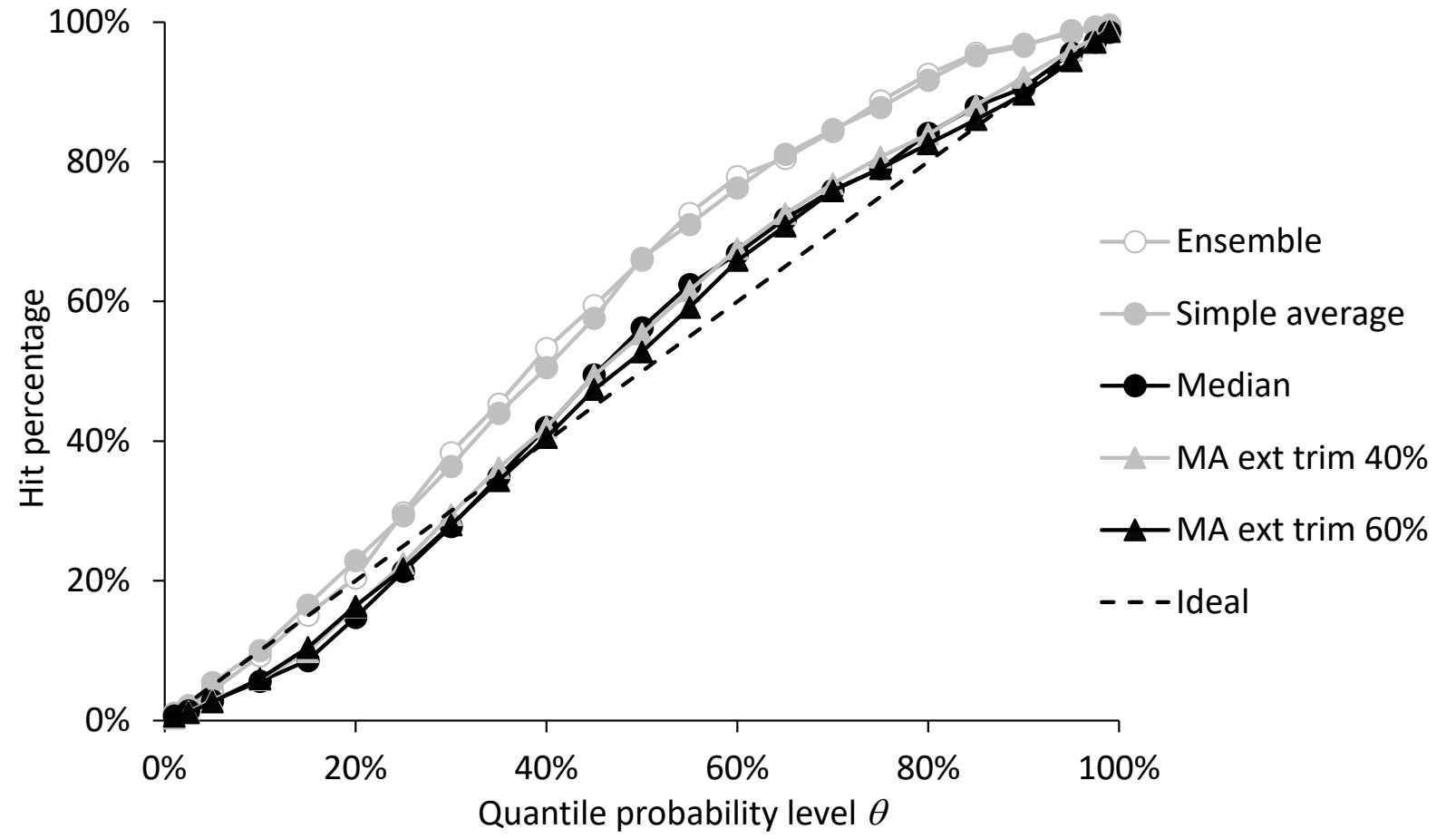

Figure 10: Calibration of distributional forecasts assessed using hit percentages for the 23 quantile probability levels $\theta$ for the medium mortality series.

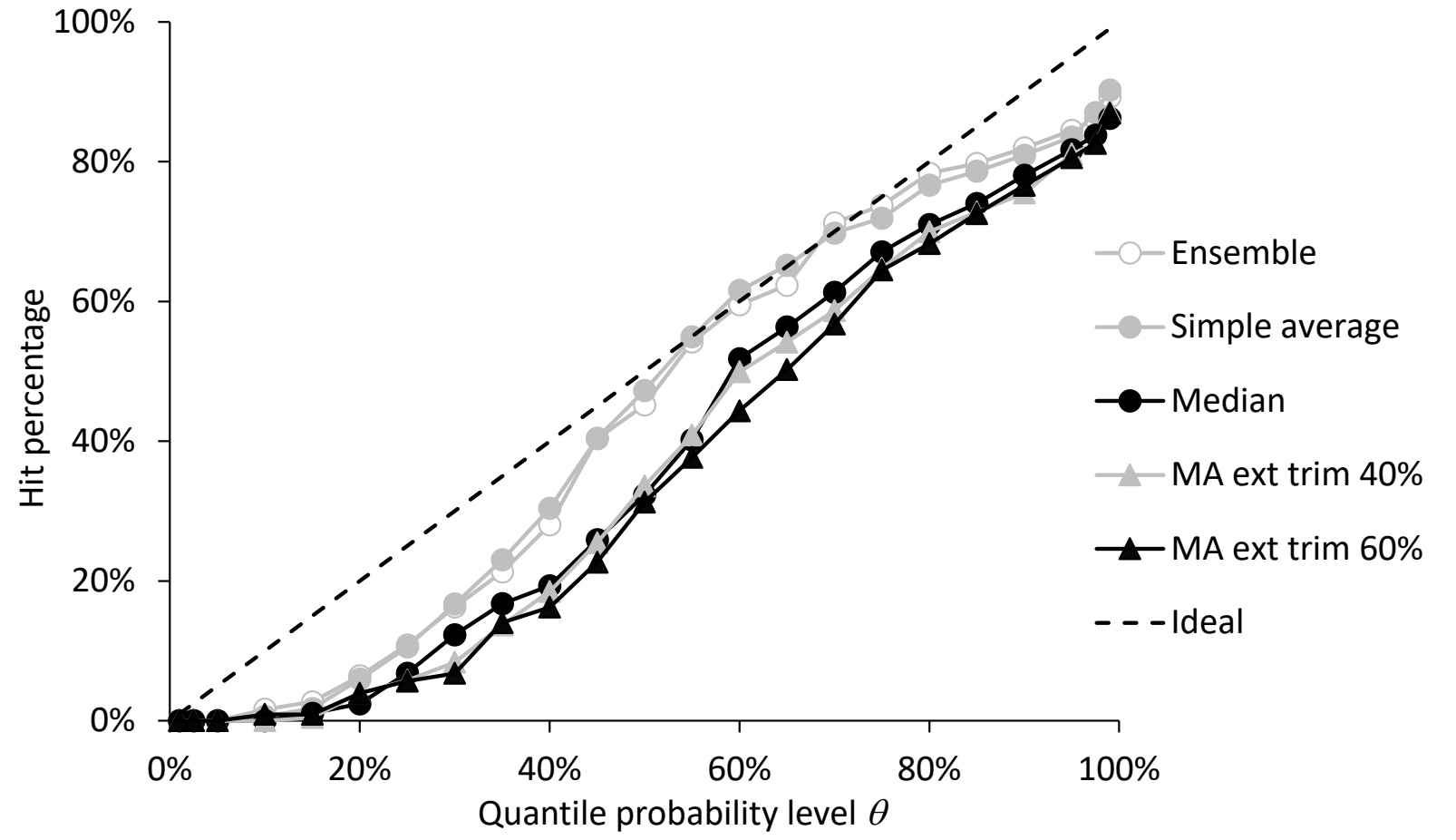

Figure 11: Calibration of distributional forecasts assessed using hit percentages for the 23 quantile probability levels $\theta$ for the high mortality series.

Table 5: Mean of the CRPS for the distributional forecasts. Aggregation methods applied to three different sets of individual forecasts: all, compartmental models only, and non-compartmental models.

| | Low | | | Medium | | | High | | |
|---|---|---|---|---|---|---|---|---|---|
| | All | Comp | Non-Comp | All | Comp | Non-Comp | All | Comp | Non-Comp |
| Ensemble | 129 | NA | NA | 737 | NA | NA | 8755 | NA | NA |
| Simple average | 129 | 159 | 125 | 758 | 848 | 900 | 7988 | 8627 | 9266 |
| Median | 94 | 109 | 112 | 657 | 718 | 807 | 8531 | 8588 | 9436 |
| CA ext trim 20% | 117 | 153 | 124 | 693 | 816 | 895 | 8309 | 8664 | 9326 |
| CA ext trim 40% | 98 | 118 | 116 | 660 | 719 | 787 | 8365 | 8648 | 9404 |
| CA ext trim 60% | 94 | 113 | 110 | 659 | 712 | 796 | 8456 | 8755 | 9189 |
| CA ext trim 80% | 92 | 109 | 111 | 658 | 714 | 804 | 8492 | 8567 | 9350 |
| CA int trim 20% | 150 | 183 | 143 | 833 | 937 | 1003 | 8138 | 8934 | 10067 |
| CA int trim 40% | 181 | 230 | 151 | 956 | 1157 | 1141 | 8368 | 9403 | 10232 |
| CA int trim 60% | 246 | 263 | 186 | 1302 | 1290 | 1552 | 9452 | 10246 | 11758 |
| CA int trim 80% | 332 | 279 | 186 | 1973 | 1382 | 1564 | 12323 | 10396 | 11781 |
| MA ext trim 20% | 117 | 153 | 124 | 694 | 816 | 896 | 8293 | 8638 | 9305 |
| MA ext trim 40% | 102 | 129 | 116 | 668 | 763 | 793 | 8292 | 8636 | 9365 |
| MA ext trim 60% | 97 | 126 | 111 | 662 | 759 | 794 | 8332 | 8799 | 9080 |
| MA ext trim 80% | 98 | 126 | 114 | 676 | 767 | 817 | 8461 | 8563 | 9344 |
| MA int trim 20% | 148 | 180 | 145 | 833 | 939 | 1033 | 8207 | 8974 | 10239 |
| MA int trim 40% | 177 | 221 | 154 | 958 | 1151 | 1179 | 8535 | 9642 | 10532 |
| MA int trim 60% | 233 | 254 | 190 | 1293 | 1305 | 1587 | 9880 | 10606 | 12180 |
| MA int trim 80% | 320 | 270 | 190 | 1985 | 1409 | 1600 | 13045 | 10926 | 12229 |

Note: The unit of CRPS is deaths. Lower CRPS values are better. NA indicates not available.

In Section 2, we discussed the different methods used by the individual forecasting teams. We described how compartmental models were used by approximately half the teams, and we illustrated this in Figure 4. We were curious to see how the aggregation methods would perform if they were applied to only the teams using compartmental models. Given their widespread use by epidemiologists, one might surmise that aggregating only these models would be adequate, and that an aggregation using only the other types of models would deliver poor results. We investigate these issues in Table 5, where we report CRPS results produced by applying the aggregation methods to the following three different sets of the individual forecasting teams: all the teams, the teams using compartmental models, and the teams not using compartmental models. For the medium and high mortality series, Table 5 shows that aggregating only compartmental models is better than aggregating only the other types of models. For the low mortality series, this is not the case. However, a clear result for each of the three groupings of mortality series is that, for the better performing methods, the greatest accuracy was achieved by aggregating all available methods.

## 5. Summary and Concluding Comments

We have provided an empirical comparison of aggregation methods for interval and distributional forecasts of cumulative mortality due to COVID-19. The forecasts were produced by teams using a variety of approaches, including compartmental and statistical models. Aggregation provides a pragmatic way to synthesize the diverse information underlying these models. The simple average is a natural benchmark, but it is also worth considering the median, as well as trimming methods, which enable robust estimation and adjustment for the case where forecasters tend to be under- or overconfident.

Given that we are currently at a relatively early stage of the pandemic, for each mortality time series, we do not have a record of historical accuracy for the same set of past periods for all forecasting teams, as we discussed in Section 2, in relation to Figure 5. Indeed, for new participants, we have no record of past accuracy. This influenced the aggregation methods that we chose to consider, as it is not clear how to optimize the weights within some form of weighted forecast combination. However, we should acknowledge that the trimming methods do require the choice of trimming parameter. This could be optimized, either for each mortality series, or by recording accuracy averaged across a group of series. An appeal of the simple average and median is that they require no parameter optimization, and in view of our empirical results, we recommend the use of the median for the low and medium mortality series, and the simple average for the high mortality series.

There is potential to improve the contribution of epidemiological models to the policy making process in dealing with the COVID-19 pandemic by considering aggregated predictions, as they offer several potential advantages, including greater accuracy, compared to predictions from single models.

## Acknowledgements

We are very grateful to all the forecasting groups who generously made their forecasts available on the Reich Lab COVID-19 Forecast Hub, and to Nicholas Reich and his team, for acting as curators for the Hub, and for providing such convenient access to the data, along with useful supplementary information. We would also like to thank Nia Roberts for clarifying our understanding of the licence terms for the forecast data.

## Supplementary Information

The terms and conditions of the forecasts that were analysed are recorded in the forecasting groups' supplementary files on the Reich Lab COVID-19 Forecast GitHub website: https://github.com/reichlab/COVID19-forecast-hub/tree/master/data-processed.
A number of the forecasting teams released their data under one of the following licences:
https://creativecommons.org/licenses/by/4.0/,
https://creativecommons.org/licenses/by-nc/4.0/,
https://creativecommons.org/licenses/by-nc-sa/4.0/

Details of forecast models, ordered by model name

| **Contributors** | **Model (short name)** | **Model description*** | **Access information (including citations)** |
|---|---|---|---|
| Wattanachit N, Ray EL, Reich N | COVID hub-ensemble | An ensemble, or model average, of submitted forecasts to the COVID-19 Forecast Hub. | https://github.com/reichlab/covid19-forecast-hub/tree/master/data-processed/COVIDhub-ensemble |
| *COMPARTMENTAL MODELS* | | | |
| Tomar V, Jain C | Auquan-SEIR | Modified SEIR model with compartments for reported and unreported infections. Non-linear mixed effects curve-fitting. | https://github.com/reichlab/covid19-forecast-hub/tree/master/data-processed/Auquan-SEIR |
| Carlson E, Henderson M, Kelly C, Kofman I, Zhang X | CovidActNow-SEIR_CAN | SEIR model forecasts of cumulative deaths, incident deaths, incident hospitalizations by fitting predicted cases, deaths, and hospitalizations to the observations. | https://github.com/reichlab/covid19-forecast-hub/tree/master/data-processed/CovidActNow-SEIR_CAN |
| Li ML, Bouardi HT, Lami OS, Trikalinos TA, Trichakis NK, Bertsimas D | CovidAnalytics-DELPHI | SEIR model augmented with underdetection and interventions. Projections account for reopening and assume interventions would be re-enacted if cases continue to climb. | https://github.com/reichlab/covid19-forecast-hub/tree/master/data-processed/CovidAnalytics-DELPHI |
| Chhatwal J, Ayer T, Linas B, Dalgic O, Mueller P, Adee M, Ladd MA, Xiao J | Covid19Sim-Simulator | An interactive tool that uses a validated SEIR compartment model. | https://github.com/reichlab/covid19-forecast-hub/tree/master/data-processed/Covid19Sim-Simulator |
| (Mass General Hospital, Harvard Medical School, Georgia Tech and Boston Medical Centre) | | | |
| Pei S, Yamana T, Kandula S, Yang W, Galanti M, Shaman J | CU-select | Metapopulation county-level SEIR model for projecting future COVID-19 incidence and deaths. This forecast is the scenario we believe to be most plausible given the current setting. | https://github.com/reichlab/covid19-forecast-hub/tree/master/data-processed/CU-select |
| Lemaitre JC, Bi Q , Hulse JD, Grabowski MK, Grantz KH, Kaminsky J, Lauer SA, Lee EC, Meredith HR, Perez-Saez J, Truelove SA, Keegan LT, Kaminsky K, Shah S, Wills J, Aquilanti P-Y, Raman K, Subramaniyan A, Thursam G, Tran A | JHU_IDD-CovidSP | County-level metapopulation model with commuting and stochastic SEIR disease dynamics with social-distancing indicators. | https://github.com/reichlab/covid19-forecast-hub/tree/master/data-processed/JHU_IDD-CovidSP |
| Baek J, Farias V, Georgescu A, Levi R, Sinha D, Wilde J, Zheng A | MITCovAlliance-SIR | SIR model trained on public heath regions. SIR parameters are functions of static demographic and time-varying mobility features. Two-stage approach that first learns magnitude of peak infections. | https://github.com/reichlab/covid19-forecast-hub/tree/master/data-processed/MITCovAlliance-SIR |
| Espana G, Oidtman R, Cavany S, Costello A, Wieler A, Lerch A, Barbera C, Poterek M, Tran Q, Moore S, Perkins A | NotreDame-Mobility | Ensemble of nine models that are identical except that they are driven by different mobility indices from Apple and Google. The model underlying each is a deterministic, SEIR-like model. | https://github.com/reichlab/covid19-forecast-hub/tree/master/data-processed/NotreDame-mobility |
| Koyluoglu U, Milliken J | OliverWyman-Navigator | Forecasts and scenario analysis for Detected and Undetected cases and death counts following a compartmental formulation with non-stationary transition rates. | https://github.com/reichlab/covid19-forecast-hub/tree/master/data-processed/OliverWyman-Navigator |
| Turtle J, Ben-Nun M, Riley P | PSI-DRAFT | A stochastic/deterministic, single-population SEIRX model that stratifies by both age distribution and disease severity and includes generic intervention fitting. | https://github.com/reichlab/covid19-forecast-hub/tree/master/data-processed/PSI-DRAFT |
| Shi Y, Shah T, Ban X | RPI-UW-Mob_Collision | A mobility-informed simplified SIR model motivated by collision theory. | https://github.com/reichlab/covid19-forecast-hub/tree/master/data-processed/RPI-UW-Mob-Collision |
| Snyder TL, Wilson DD | SWC-TerminusCM | Mechanistic compartmental model using disease parameter estimates from literature. It | https://github.com/reichlab/covid19-forecast-hub/tree/master/data-processed/SWC-TerminusCM |
| | | uses Bayesian inference to predict the most likely model parameters. | |
| Cobey S, Arevalo P, Baskerville E, Carran S, Gostic K, McGough L, Ranjeva S, Wen F | UChicago-COVIDIL | Compartmental, age-structured SEIR model that infers past SARS-CoV-2 transmission rates and forecasts mortality under current and hypothetical public health interventions. | https://github.com/reichlab/covid19-forecast-hub/tree/master/data-processed/UChicago-CovidIL |
| Gu Q, Xu P, Chen J, Wang L, Zou D, Zhang W | UCLA-SuEIR | Variant of the SEIR model considering both untested and unreported cases. The model considers reopening and assumes susceptible population will increase after the reopen. | https://github.com/reichlab/covid19-forecast-hub/tree/master/data-processed/UCLA-SuEIR |
| Chen YQ, Zhao Y, Guo L | UCM-MESALab-FoGSEIR | FoGSEIR model is a modification of integer order SEIR model considering fractional integrals. The model considers the age structure and reopening intervention to minimize infections and deaths. | https://github.com/reichlab/covid19-forecast-hub/tree/master/data-processed/UCM_MESALab-FoGSEIR |
| Sheldon D, Gibson G, Reich N | UMass-M echBayes | Bayesian compartmental model with observations on cumulative case counts and cumulative deaths. Model is fit independently to each state. Model includes observation noise and a case detection rate. | https://github.com/reichlab/covid19-forecast-hub/tree/master/data-processed/UMass-MechBayes |
| Mayo ML, Rowland MA, Parno MD, Detwiller ID, Farthing MW, England WP George GE | USACE-ERDC_SEIR | The ERDC SEIR model makes predictions of several variables (e.g., reported new/cumulative cases per day, etc.). Model parameters are estimated using historical data using Bayesian inference. | https://github.com/reichlab/covid19-forecast-hub/tree/master/data-processed/USACE-ERDC_SEIR |
| Gu Y | YYG-ParamSearch | Based on the SEIR model with hyperparameter optimization to make daily projections regarding COVID-19 infections and deaths in 50 US states. The model accounts for state reopenings and its effects on infections and deaths. | https://github.com/reichlab/covid19-forecast-hub/tree/master/data-processed/YYG-ParamSearch |
| *OTHER MODELS* | | | |
| O'Dea E | CEID-Walk | A random walk model with drift. A least squares line is fitted to the tail observations of a target time series to estimate the drift and step variance of a random walk model. | https://github.com/reichlab/covid19-forecast-hub/blob/master/data-processed/CEID-Walk/metadata-CEID-Walk.txt |
| Wang Y, Zeng D, Wang Q, Xie S | Columbia_UNC-SurvCon | Survival-convolution model with piece-wise transmission rates that incorporates latent | https://github.com/reichlab/covid19-forecast-hub/tree/master/data- |
| | | incubation period and provides time-varying effective reproductive number. | processed/Columbia_UNC-SurvCon |
| Ray EL, Tibshirani R | COVIDhub-baseline | Baseline prediction model. | https://github.com/reichlab/covid19-forecast-hub/tree/master/data-processed/COVIDhub-baseline |
| Kalantari R | DDS-NBDS | Jointly modeling daily deaths and cases using a negative binomial distribution based nonparametric Bayesian generalized linear dynamical system. | https://github.com/reichlab/covid19-forecast-hub/tree/master/data-processed/DDS-NBDS |
| Sherratt K, Bosse N, Abbott S, Hellewell J, Meakin S,Munday J, Funk S, | epiforecasts-ensemble1 | A deaths forecast using the renewal equation and time-series forecasts of the time-varying reproduction number. | https://github.com/reichlab/covid19-forecast-hub/tree/master/data-processed/epiforecasts-ensemble1 |
| Keskinocak P, Aglar BEO, Baxter A, Asplund J, Serban N | GT_CHHS-COVID19 | Agent-based simulation model to project COVID19 infection spread. | https://github.com/reichlab/covid19-forecast-hub/tree/master/data-processed/GT_CHHS-COVID19 |
| Prakash BA, Rodriguez A, Cui J, Tabassum A, Adhikari B, Sun J, Xiao D, Qiang C (Georgia Institute of Technology, Virginia Tech, University of Iowa, University of Illinois, IQVIA) | GT-DeepCOVID | Data-driven approach based on deep learning for forecasting mortality and hospitalizations using syndromic, clinical, demographic, mobility and point-of-care data. | https://github.com/reichlab/covid19-forecast-hub/tree/master/data-processed/GT-DeepCOVID |
| Murry C and the IHME-CurveFitTeam | IHME-CurveFit | Non-linear mixed effects curve-fitting. This model makes predictions about the future that are dependent on the assumption that current interventions continue. | https://github.com/reichlab/covid19-forecast-hub/tree/master/data-processed/IHME-CurveFit |
| Wang L, Wang G, Gao L, Li X, Yu S, Kim M, Wang Y, Gu Z | IowaStateLW-STEM | A nonparametric space-time disease transmission model. The projections assume that the data used is reliable, the future will continue to follow the current pattern, and current interventions will remain the same till the end of forecasting period. | https://github.com/reichlab/covid19-forecast-hub/tree/master/data-processed/IowaStateLW-STEM |
| Karlem D | Karlen-pypm | python Population Modeller | https://github.com/reichlab/covid19-forecast-hub/tree/master/data-processed/Karlen-pypm |
| Osthus D, Del Valle S, Manore C, Weaver B, Castro L, Shelley S, Smith M, Spencer J, Fairchild G, Travis Pitts T, Gerts D, Dauelsberg L, Daughton A, Gorris M, Hornbein B, Israel D, Parikh N, Shutt D, Ziemann A | LANL-GrowthRate | Statistical dynamical growth model accounting for population susceptibility. Makes predictions about the future, unconditional on particular intervention strategies. | https://github.com/reichlab/covid19-forecast-hub/tree/master/data-processed/LANL-GrowthRate |
| Vespignani A, Chinazzi M, Davis JT, Mu K, Pastore y Piontti A, Samay N, Xiong X, Halloran ME, Longini IM, Dean NE, Viboud C, Sun K, Litvinova M, Gioannini C, Rossi L, Ajelli M | MOBS-GLEAM_COVID§ | Metapopulation, age structured SLIR model. Superimposed on the worldwide population and mobility layers is an agent-based epidemic model that defines the infection and population dynamics. Makes predictions about the future that are dependent on the assumption that current interventions continue. | https://github.com/reichlab/covid19-forecast-hub/tree/master/data-processed/MOBS-GLEAM_COVID |
| Espana G, Oidtman R, Cavany S, Costello A, Wieler A, Lerch A, Barbera C, Poterek M, Tran Q, Moore S, Perkins A | NotreDame-FRED | Agent-based model developed for influenza with parameters modified to represent the natural history of COVID-19 | https://github.com/reichlab/covid19-forecast-hub/tree/master/data-processed/NotreDame-FRED |
| Walraven R | RobertWalraven-ESG | Multiple skewed gaussian distribution peaks fitted to raw data. | https://github.com/reichlab/covid19-forecast-hub/tree/master/data-processed/RobertWalraven-ESG |
| Bieggel H, Lega J | UA-EpiCovDA† | SIR mechanistic model with data assimilation. EpiCovDA is an extension of the EpiGro model. Model parameters are fit to Covid-19 data using a variational data assimilation method. A prior distribution of the parameters is estimated by fitting an SIR Incidence-Cumulative Cases curve to data from states that had at least 1000 cases by 04/01/2020. | https://github.com/reichlab/covid19-forecast-hub/tree/master/data-processed/UA-EpiCovDA |
| Corsetti S, Schwarz T | UMich-RidgeTfReg | Nation-level model of confirmed cases and deaths based on ridge regression.<br>No assumptions made about social distancing. | https://github.com/reichlab/covid19-forecast-hub/tree/master/data-processed/UMich-RidgeTfReg |
| Woody S, et al. at the University of Texas | UT-Mobility | This model makes predictions assuming that social distancing patterns, as measured by anonymized mobile-phone GPS traces, remain constant in the future. Only models *first-wave deaths*. | https://github.com/reichlab/covid19-forecast-hub/tree/master/data-processed/UT-Mobility |

* Based on information recorded on the GIT Hub
§ Classed as other model although it has a compartmental structure (SLIR), as it also has agent-based elements. This is a hybrid model.
† Classed as other model as it is an extension to a non-linear growth model with a prior distribution SIR curve fitted. Therefore, it is not a compartmental.